\documentclass[aps,prl,twocolumn,showpacs,superscriptaddress,groupedaddress,superscriptaddress]{revtex4-1}  
\usepackage{graphicx}  
\usepackage{dcolumn}   
\usepackage{bm}        
\usepackage{amssymb}   
\usepackage[usenames]{color}

\begin{document}

\title{
Interstitial Channels that Control Band Gaps and Effective Masses in Tetrahedrally Bonded Semiconductors
}

\author{Yu-ichiro Matsushita}
\affiliation{Max-Planck Institute of Microstructure Physics, Weinberg 2, D-06120 Halle, Germany}
\affiliation{Department of Applied Physics, The University of Tokyo, 
Tokyo 113-8656, Japan}
\author{Atsushi Oshiyama}
\affiliation{Department of Applied Physics, The University of Tokyo, 
Tokyo 113-8656, Japan}

\date{\today}

\begin{abstract}
We find that electron states at the bottom of the conduction bands of covalent semiconductors are distributed mainly in the interstitial channels and that this {\em floating} nature leads to the band-gap variation and the anisotropic effective masses in various polytypes of SiC. We find that the channel length, rather than the hexagonality prevailed in the past, is the decisive factor for the band-gap variation in the polytypes. We also find that the {\em floating nature} causes two-dimensional electron and hole systems at the interface of different SiC polytypes and even one-dimensional channels near the inclined SiC surface.
\end{abstract}

\pacs{}
\maketitle

\begin{figure}
\includegraphics[width=0.9\linewidth]{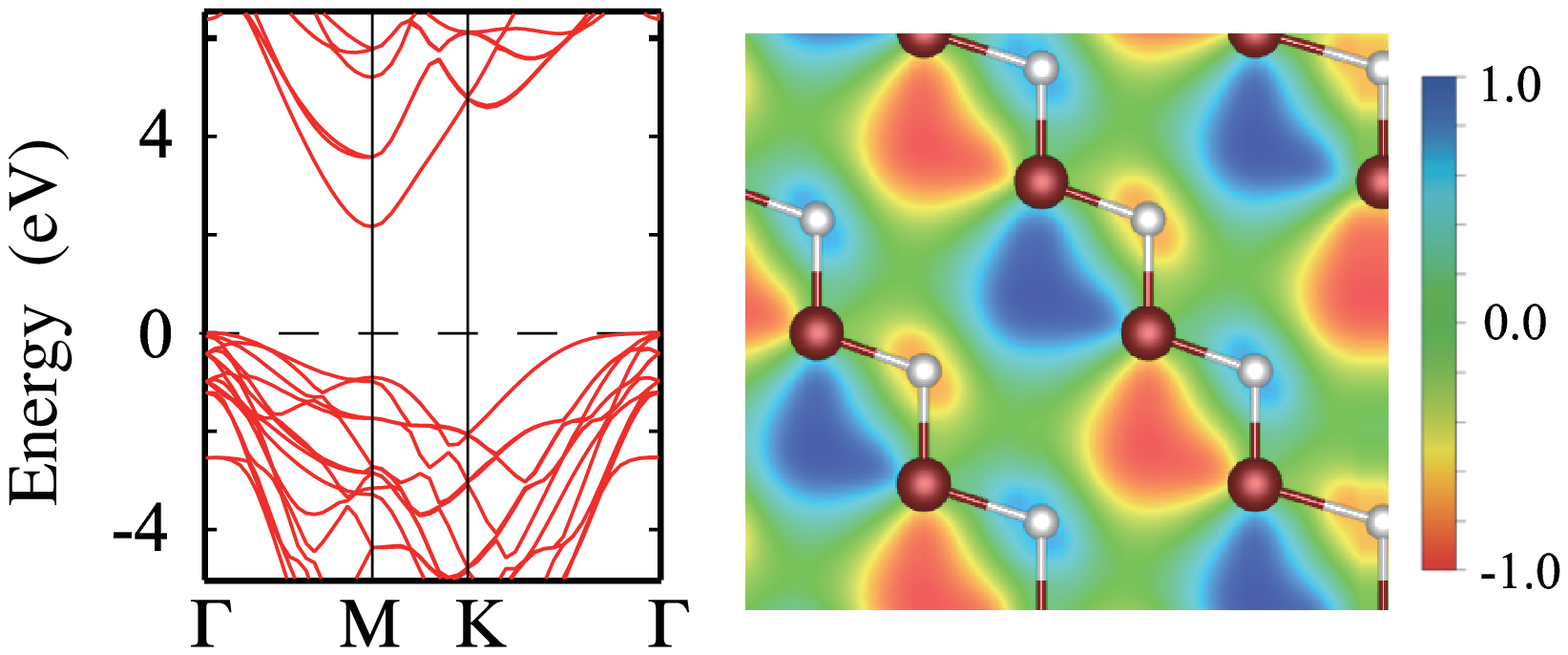}
\caption{(Color online). 
Energy bands and the contour plot of the Kohn-Sham orbital of the conduction-band minimum at $M$ in $3C$-SiC shown on $(0\bar{1}1)$ plane. The orbital is distributed along the [110] channel. The calculation has been done with the hybrid exchange-correlation functional, HSE \cite{hse,matsushita}. 
White and burgundy balls depict C and Si atoms, respectively.
Simple localized-orbital basis sets are incapable of describing the floating nature of the conduction-band states \cite{Matsushita_floating,matsushita_th}
}
\label{float}
\end{figure}

Most semiconductors, elemental or compound, have the four-fold coordinated tetrahedral structure caused by the hybridization of atomic orbitals. It is written in textbooks \cite{phillips} that the resultant hybridized $sp^3$ bonding orbitals constitute valence bands, whereas the anti-bonding counter parts do conduction bands. This is not necessarily true, however: We have recently found that the wave-functions of the conduction-band minima (CBM) of the semiconductors are distributed not near atomic sites but in the interstitial channels \cite{Matsushita_floating}, as shown in Fig.~\ref{float}. The wave-functions {\em float} in the internal space, i.e., the channels, inherent to the $sp^3$-bonded materials. 

Another structural characteristic in the semiconductor is the stacking of atomic bilayers along the bond axis direction such as AB (wurtzite) or ABC (diamond or zincblende). The different stacking sequence leads to the different polytype \cite{Dissertation10} generally labeled by the periodicity of the sequence $n$ and its symmetry, hexagonal ($H$) or cubic ($C$), as in 2$H$(AB), 3$C$(ABC), 4$H$(ABCB) and so on. These differences in the stacking sequence have been assumed to be minor in the electronic properties. However, the sequence determines the lengths and the directions of the interstitial channels, hereby affecting the shapes of the wave-functions of CBMs. The internal space overlooked in the past may be closely related to the electronic properties of the semiconductors, that we discuss in this Letter.

Silicon carbide (SiC) is a promising material in power electronics due to its superior properties which are suitable to the operations under harsh environment \cite{Matsunami}. From science viewpoints, SiC is a manifestation of the polytypes explained above: Dozens of polytypes of SiC are observed and the band gaps vary by 40 \%, from 2.3 eV in 3$C$ to 3.3 eV in 2$H$ despite that the structures are locally identical to each other in the polytypes \cite{Harris}. This mysterious band-gap variation has been discussed in terms of an empirical quantity, hexagonality \cite{Choyke}, for a half century: A bilayer sandwiched by the two same stacking indexes, as in 2H structure, is called a hexagonal layer and the ratio of the hexagonal layers in whole stacking sequence is called hexagonality; the band-gap variation in the polytypes is argued to be linear with respect to the hexagonality. Yet, the linearity is not satisfactory (see below) and moreover the underlying physics is totally lacking.

In this Letter, we find, on the basis of the density-functional calculations \cite{cal}, that the extent of the internal space, i.e., the length of the interstitial channel, in covalent semiconductors is decisive in the nano-scale shapes of the wave-functions of the CBM and hereby explains the mysterious variation of the band gap in SiC polytypes. We also find that the observed anisotropy of the effective masses in SiC, and the pressure dependence of the band gaps generally observed in most semiconductors, are naturally explained in terms of the channel length. Further, we find that the stacking control dramatically modifies the electronic properties, leading to generation of low-dimensional electron and hole systems in three-dimensional SiC.

The sequence of the atomic bilayers determines the length and direction of the interstitial channels: e.g., in the 3$C$ polytype the channel along $\langle 110 \rangle$ extends infinitely, whereas in the 6$H$ polytype the cannel along $\langle 2\bar{2}01 \rangle$ has a finite length of 7$a_0$ / 2 $\sqrt{2}$ ($a_0$: lattice constant). To examine the relation between the extent of the internal space and the band gap, we consider 24 representative stacking sequences in 3$C$ and $nH$ (2 $\le n \le$ 12) polytypes. Details of the 24 polytypes are listed in Supplement Material \cite{SM}. In the 2$H$, 3$C$, 4$H$ and 5$H$ polytypes, the sequence of the bilayer stacking is unique. In the $6H$, $8H$, $10H$, and $12H$ polytypes, there are 2, 6, 18, and 58 possibilities in the stacking sequence, respectively \cite{Iglesias}. The possible values of the hexagonality in the $10H$ polytype are 20, 40, 60, and 80\%, whereas those in the $12H$ polytype are 16.7, 33.3, 50, 66.7, and 83.3\%. Our 24 representatives include all the possible hexagonality in the the $6H$, $8H$, $10H$, and $12H$ polytypes \cite{Kobayashi}. 

\begin{figure}
\includegraphics[width=1.0\linewidth]{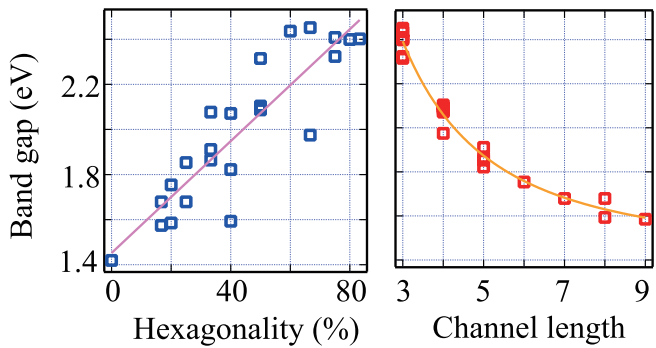}
\caption{(Color online). 
Band gaps for 24 representative SiC polytypes calculated in GGA as a function of the hexagonality (left panel) and a function of the channel length (right panel). 
In each panel, the fitting function (see text) is also shown. The corresponding variance is 355 meV for the left (hexagonality) and 85 meV for the right (channel length). 
}
\label{channel_length}
\end{figure}

Calculated band gaps for these representative polytypes of SiC are plotted as a function of either the hexagonality or the channel length in Fig.\ref{channel_length}. Here the channel length is defined as the number of bilayers along the longest interstitial channels. The left panel in Fig.~\ref{channel_length} shows a positive correlation between the band gap and the hexagonality. 
However, the linearity between the two is poor: The best fitted function we have obtained is $\epsilon_g=1.425+0.0124 h$ in eV (h: hexagonality) with the large variance of 355 meV. 
On the other hand, the band gap as a function of the channel length shows monotonic decrease, indicating that the channel length is a proper quantity to describe the band gap. We have indeed found that the calculated band gaps $\varepsilon_{g}$ are nicely fitted to a single function of the channel length $l$ as,  
\begin{equation}
\varepsilon_g = 1.425 + \frac{17.63}{( l + 1.268)^2} \ ,
\label{fitting}
\end{equation}
in eV with the variance of 85 meV. 

As stated above, the wavefunction of the CBM {\em floats} in the interstitial channel the length $l$ of which is determined by the way of stacking of atomic bilayers. Hence, the CBM state is regarded as being confined in the one-dimensional quantum tube with the length $l$. The energy level $\varepsilon_l$ thus confined is given by \cite{matsushita_th}, 
\begin{equation}
\varepsilon_l = \varepsilon_{3C} + \frac{\pi^2 \hbar^2}{2m^* (l + \Delta)^2} \ ,
\label{quantum_well}
\end{equation}
where $\varepsilon_{3C}$ is the energy level of 3$C$-SiC which has the infinite channel length and $m^*$ is the effective mass along the channel direction. 
The $\Delta$ in the second term represent the spill of the wave function from the quantum tube with the length $l$. 
Since the valence-band top has a character of the bonding $sp^3$ orbitals common to all the polytypes \cite{Matsushita_floating,matsushita_th}, the variation in (\ref{quantum_well}) corresponds to that in the band gap in the polytypes expressed in (\ref{fitting}). Our GGA calculation indeed provides $\varepsilon_{3C}$ = 1.419 eV, showing agreement with the first term in (\ref{fitting}). By further comparing the second terms in (\ref{fitting}) and (\ref{quantum_well}), we obtain the effective mass of $m^* = 0.326 m_0$ ($m_0$: bare electron mass) which shows agreement with the experimental value of $m^* = 0.363 m_0$ \cite{3C-SiC}. 
The factor 1.268 in unit of the bilayer length in the denominator of the fitting function (\ref{fitting}) indicates that 
the confinement is imperfect and the wavefunction spills from the interstitial tube by about a single bilayer. 

\begin{figure}
\includegraphics[width=1.0\linewidth]{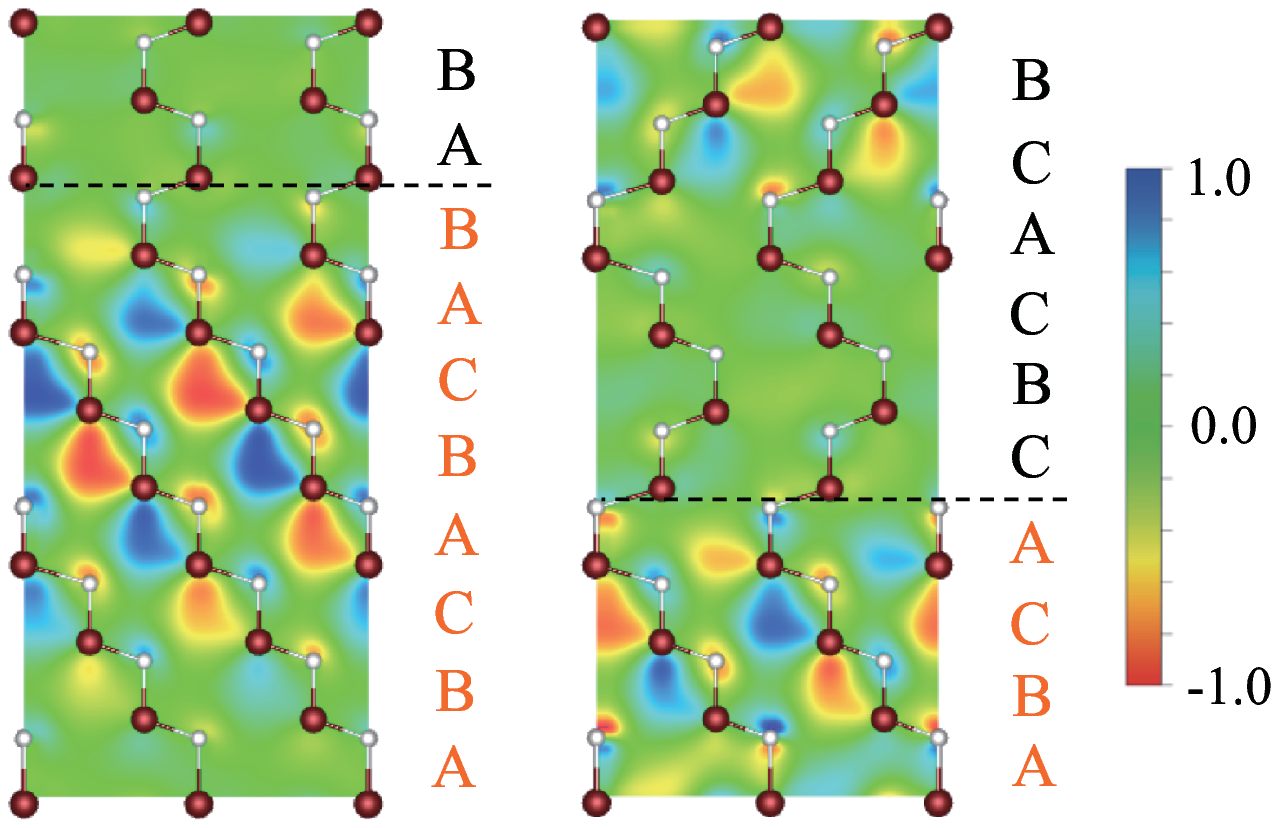}
\caption{(Color online). 
Contour plots on ($11\bar{2}0$) plane of the Kohn-Sham (KS) orbitals of the conduction band minimum at $M$ for $10H$-SiC with the ABCABCABAB stacking (left panel) and with the ABCACBCACB stacking (right panel). The value for each contour color is relative to the corresponding maximum absolute value. White and burgundy balls depict C and Si atoms, respectively. The broken lines represent the interface 
where the longest channel in the cubic region is blocked by the hexagonal region. 
}
\label{distribution}
\end{figure}

We have now clarified that the channel length rather than the hexagonality is the principal quantity to determine the band-gap variation. This situation becomes visible by examining the wavefunction of the CBM. Figure \ref{distribution} shows the Kohn-Sham orbital of the CBM for the two 10$H$-SiC polytypes where the stacking sequences are ABCABCABAB and ABCACBCACB, respectively. The hexagonality of the two polytypes is identical, i.e., 40 \%. However, the calculated band gap is 1.59 eV for the former and 2.07 eV for the latter. This difference in the gap beautifully corresponds to that in their channel lengths, 8 and 4, respectively \cite{SM}. The wavefunction of CBM {\em floats} along the $\langle 10\ \bar{10}\ 0 3 \rangle$ for both polytypes as is shown in Fig.~\ref{distribution}. Yet the length of the channel and consequently the extension of the wavefunction is substantially longer for the former polytype, leading to the narrower band gap. 

\begin{figure}
\includegraphics[width=1.0\linewidth]{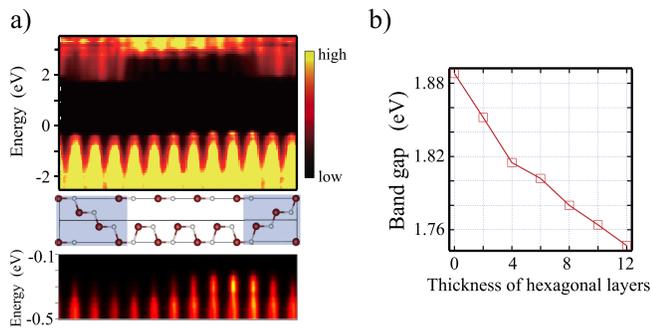}
\caption{(Color online). 
(a) Local density of states, $D(\varepsilon, z)$, near the band gap for the $12H$-SiC (ABCACACACACB) with its side view of the atomic structure. 
The shaded region in the side view depicts the cubic stacking region. The lower panel is  
the extension of $D(\varepsilon, z)$ near the valence-band top. The ordinate $\varepsilon$ is the electronic energy, where the Fermi energy is set to be 0, and the abscissa $z$ is the coordinate along the stacking direction. The value of $D ( \varepsilon, z)$ is represented by the color code (Yellow: high, black: low). 
The black region corresponds to the energy gap. 
(b) Band gap variation as a function of the thickness of hexagonal-stacking region of SiC polytypes with the thickness of the cubic stacking region fixed with 5 bilayers.  
}
\label{SP}
\end{figure}

The calculated band gaps in the right panel of Fig. \ref{channel_length} show small but sizable variance from the fitting function described above. This variance is a consequence of spontaneous polarization (SP) in the region of the hexagonally stacked bilayers (hexagonal stacking). Let us consider the polytypes with the 4-bilayer channel length in Fig.~\ref{channel_length}: A $12H$ polytype whose stacking sequence is ABCACACACACB shows narrower band gap by 0.1 eV than those of other polytypes with $l$ = 4. Figure \ref{SP} (a) shows calculated local density of states (LDOS) near the energy gap for this $12H$-SiC. The spiky contrast below the energy gap (black region) manifests atomic positions along the stacking direction. It is clearly shown that the conduction bands in the region with the cubic stacking (ABC) are located at lower positions in energy than in the hexagonal stacking region. More importantly, SP takes place in the hexagonally stacked region due to the lack of inversion symmetry and renders the band lineup slanted in real space along the stacking direction. Further the counter polarization in the cubic region makes it slanted in the reverse direction as in Fig.~\ref{SP} (a). We have found that the slanted band lineup causes downward (upward) shift of the conduction (valence) band edge and the band gap becomes narrower. 

We have indeed calculated the band-gap variation by increasing the thickness of the hexagonal bilayers with the thickness of the cubic region fixed at 5 bilayer [Fig.~\ref{SP} (b)]. The calculated band gaps decreases monotonically. This is a consequence from the enhanced band-slanting induced by the polarization. The estimated band-gap decrease by adding a single bilayer in the hexagonally stacked region is 10 meV. By using this quantity, the estimated band gap of the $12H$-SiC (ABCACACACACB) {\em without} SP is 2.07 eV, which is just on the fitting function in Fig.\ref{channel_length}. This finding opens a possibility of the band-gap tuning by changing the thickness of the hexagonally stacked region.

We have revealed that the floating nature of the CBM causes the band-offset at the interface between the cubic-stacking and the hexagonal-stacking regions. The SP in the hexagonal region combined with the counter polarization in the cubic region generates two-dimensional electron and hole gases at the interface. It may be evident from LDOS slanted in real space, as is shown in Fig.\ref{SP} (a). It is further quantified by calculating the effective masses of the CBM and the valence-band maximum along the stacking direction in the polytypes: It is found that the effective mass along the direction increases to a hundred of $m_0$, whereas that in the lateral plane keeps its value of 0.67 $m_0$ for electron and 2.20 $m_0$ for hole (not shown). This finding of the carrier confinement is the generalization of the hetero-crystalline superlattice of SiC first proposed by Bechstedt and K\"ackel \cite{bechstedt} and later pursued theoretically \cite{ke,miao,iwata}. We here emphasize that the underlying physics, unrevealed in the past, is the floating nature of the CBM states controlled by the nanoshapes of the interstitial channels. The anisotropic effective mass observed in 6$H$-SiC polytype \cite{6H-SiC} is one of the fingerprints of such floating nature. 

\begin{figure}
\includegraphics[width=1.0\linewidth]{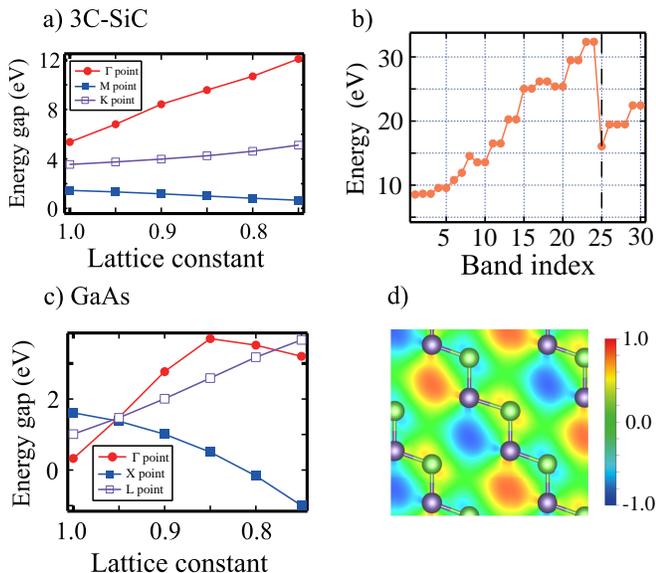}
\caption{(color online) 
Calculated CBM energy at several $k$ points as a function of reduced lattice constant $a/a_0$ ($a_0$: lattice constant without pressure) in 3C-SiC (a) and in GaAs (c). The corresponding pressures (GPa) at $a/a_0$ = 0, 0.95, 0.9, 0.85, 0.80, and 0.75 are 0 (0), 31.0 (5.9), 73.7 (12.5), 128.7 (21.3), 202.6 (33.6), and 306.5 (51.5) in SiC (GaAs). (b) The kinetic-energy contribution $\epsilon_{\rm kin}=\left<\phi_i\left|-\nabla^2/2\right|\phi_i\right>$ to the orbital energy of each Kohn-Sham (KS) state at $M$ point in 3C-SiC. The abscissa represents the $i$th KS state from the valence-band bottom and the 25th state is the CBM. (d) Contour plots of the KS orbital of the CBM in GaAs at $a=0.75a_0$. The gray and green balls represent Ga and As atoms, respectively.
}
\label{pressure_dependence}
\end{figure}

Another noteworthy feature of the floating state is its pressure dependence. Fig.~\ref{pressure_dependence} (a) shows the pressure dependence of the CBM energy at several high-symmetry $k$ points. The $M$ point energy shifts downward with increasing the pressure, whereas the $\Gamma$ point energy shifts upward. The latter is easily understood by the enhancement of the bonding-antibonding splitting. The former is a consequence of the floating nature. As shown in Fig.~\ref{pressure_dependence} (b), the reduction of the kinetic-energy contribution to the orbital energy is one of the characteristics of the floating state. This is due to the extended distribution in the internal space of the floating state. When the lattice constant is reduced under the pressure, the kinetic energy generally increases but in the floating state such increase is minor. This causes the lack of the upward shift or even the downward shift at the $M$ point. The floating nature is not restricted to SiC. Figure~\ref{pressure_dependence} (c) shows the CBM energy variation in pressurized GaAs. The CBM at $M$ point, which is folded from the $X$ point in the cubic BZ, shifts downward with increasing pressure. We have actually found that the KS orbital at $M$ point has floating character in pressurized circumstances [see Fig.~\ref{pressure_dependence} (d)]. The direct- and indirect-gap transition in the pressurized GaAs well established in experiments \cite{GaAs} is a manifestation of the floating nature of the CBM states.

\begin{figure}
\includegraphics[width=1.0\linewidth]{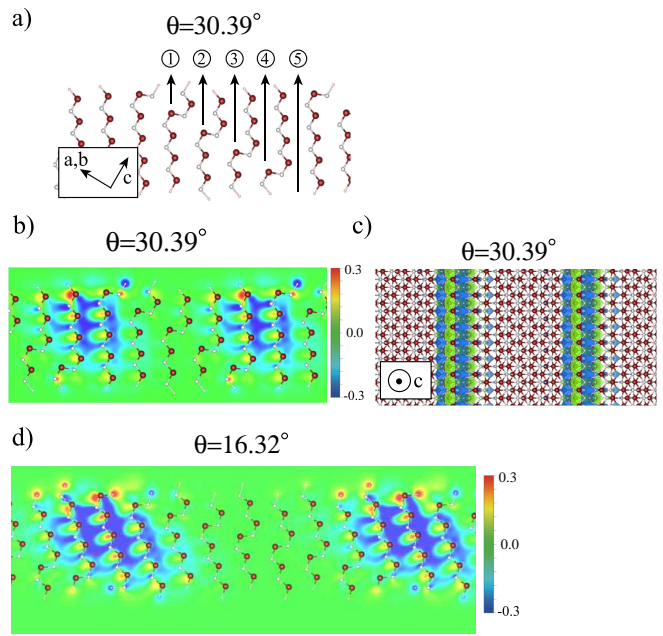}
\caption{(color online) 
One dimensional electron channel on the inclined SiC surface (xylophone channel). (a) Schematic side view of $5H$-SiC near the inclined surface (the inclined angle $\theta$ ) on $( 11\bar{2}0)$ plane. The five arrows represent channels with each number discriminating the channel length. (b) and (d): Contour plots on the $( 11\bar{2}0)$ plane of the CBM KS orbitals near the surface with the inclined angle of $\theta=30.39^\circ$ and $\theta=16.32^\circ$. (c) Top view of the CBM KS orbitals as an iso-value surface at its value of 20\% of the maximum value. Blue and green iso-value surfaces represent the positive and negative signs of the KS orbital. White and burgundy balls depict C and Si atoms, respectively.
}
\label{xylophonemodel}
\end{figure}

Nanofabrication of the semiconductor surfaces introduces further modification of wavefunctions of the floating states by controlling the internal space. Suppose, e.g.,  $5H$-SiC which has the interstitial channels along the $\langle 5\bar{5}03 \rangle$ direction with the finite length of 5 bilayers. When the surface is inclined relative to the $\langle 0001 \rangle$ direction with the angle of $\theta$, the lengths of the channels near the surface vary depending on the lateral positions like a xylophone [Fig.~\ref{xylophonemodel} (a)]. In this case, the lowest CBM state is distributed along the longest channel [Fig.\ref{xylophonemodel} (b), and (c) for $\theta = 30.39^\circ$]. It is clearly seen that one-dimensional (1D) electron channel appears near the surface. In fact, the effective mass along the 1D channel is 0.34 $m_0$ whereas the mass along the perpendicular direction is 71 $m_0$. By changing the inclined angle $\theta$, the width of the 1D channel and its separation from the adjacent channels are controlled, as is demonstrated in Fig.~\ref{xylophonemodel} (d) for the case of $\theta = 16.32^\circ$. 

We have clarified that channel structure plays important roles in understanding of the band gap variation in SiC polytypes. This variation is not limited to SiC since CBMs of most tetrahedrally-bonded materials have floating nature \cite{Matsushita_floating}. The band-gap engineering through the control of the nanoshapes of the interstitial channels comes true when syntheses of the polytypes in other materials are realized. 

In summary, we have found that the internal nanospace plays a decisive role in determining the band gaps, the effective masses and then the electronic properties of the covalent semiconductors. This is a consequence of the floating nature of the conduction-band minima where the wave-functions are distributed along the interstitial channels in the semiconductors. We have shown that the band-gap variation in various polytypes in SiC is quantitatively explained in terms of the channel length. We have found that the length and the direction of the interstitial channel are the principal quantities to determine the anisotropic effective masses, leading to the two dimensional electron and hole gases in heterocrystalline SiC and also to the one-dimensional electron channel by the surface fabrication.

This work was supported by the Grants-in-Aid for scientific research conducted by MEXT, Japan, under Contract No. 22104005. Computations were performed mainly at the Supercomputer Center in ISSP, The University of Tokyo.

\end{document}